\documentclass[aps,prl,reprint,superscriptaddress,amssymb,floatfix,longbibliography]{revtex4-1}

\usepackage{graphicx}
\usepackage{color}
\usepackage{dcolumn}
\usepackage{amsmath}
\usepackage{amssymb}
\usepackage{epstopdf}
\usepackage{braket,longtable}
\usepackage[usenames,dvipsnames,svgnames,table]{xcolor}

\usepackage[pdftex]{hyperref}
\hypersetup{colorlinks=true,linkcolor=blue,citecolor=blue,urlcolor=blue}

\newcommand{\ccag}{Ce$_2$CoAl$_7$Ge$_4$}
\newcommand{\slrr}{$T_1^{-1}$}

\begin{document}

\title{Nuclear magnetic resonance investigation of the heavy fermion system {Ce$_2$CoAl$_7$Ge$_4$}}

\author{A. P. Dioguardi}
\email[]{apdioguardi@gmail.com}
\affiliation{Los Alamos National Laboratory, Los Alamos, New Mexico 87545, USA}

\author{P. Guzman}
\affiliation{Los Alamos National Laboratory, Los Alamos, New Mexico 87545, USA}
\affiliation{Department of Physics and Astronomy, UCLA, Los Angeles, California 90095, USA}

\author{P. F. S. Rosa}
\affiliation{Los Alamos National Laboratory, Los Alamos, New Mexico 87545, USA}

\author{N. J. Ghimire}
\altaffiliation[Present Address: ]{Argonne National Laboratory, Argonne, Illinois 60439, USA}
\affiliation{Los Alamos National Laboratory, Los Alamos, New Mexico 87545, USA}

\author{S. Eley}
\affiliation{Los Alamos National Laboratory, Los Alamos, New Mexico 87545, USA}

\author{S. E. Brown}
\affiliation{Department of Physics and Astronomy, UCLA, Los Angeles, California 90095, USA}

\author{J. D. Thompson}
\author{E. D. Bauer}
\author{F. Ronning}
\affiliation{Los Alamos National Laboratory, Los Alamos, New Mexico 87545, USA}

\date{\today}

\begin{abstract}
We present nuclear magnetic resonance (NMR) and nuclear quadrupole resonance (NQR) measurements performed on single crystalline \ccag{}, a member of a recently discovered family of heavy fermion materials Ce$_2M$Al$_7$Ge$_4$ ($M$ = Co, Ir, Ni, or Pd). Previous measurements indicated a strong Kondo interaction as well as magnetic order below $T_M = 1.8$ K. Our NMR spectral measurements show that the Knight shift $K$ is proportional to the bulk magnetic susceptibility $\chi$ at high temperatures. A clear Knight shift anomaly ($K \not\propto \chi$) is observed at coherence temperatures $T^* \sim 17.5$ K for $H_0 \parallel \hat{c}$ and 10 K for $H_0 \parallel \hat{a}$ at the ${}^{59}$Co site, and $T^* \sim 12.5$ K at the ${}^{27}$Al(3) site for $H_0 \parallel \hat{a}$ characteristic of the heavy fermion nature of this compound. At high temperatures the ${}^{59}$Co NMR spin-lattice relaxation rate $T_1^{-1}$ is dominated by spin fluctuations of the 4$f$ local moments with a weak metallic background. The spin fluctuations probed by ${}^{59}$Co NMR are anisotropic and larger in the basal plane than in the $c$ direction. Furthermore, we find $(T_1TK)^{-1} \propto T^{-1/2}$ at the ${}^{59}$Co site as expected for a Kondo system for $T > T^*$ and $T> T_K$. ${}^{59}$Co NQR \slrr{} measurements at low temperatures indicate slowing down of spin fluctuations above the magnetic ordering temperature $T_M \sim 1.8$ K. A weak ferromagnetic character of fluctuations around $\mathbf{q}=0$ is evidenced by an increase of $\chi T$ versus $T$ above the magnetic ordering temperature. We also find good agreement between the observed and calculated electric field gradients at all observed sites.
\end{abstract}

\maketitle

\section{Introduction}
\label{sec:introduction}

Strong electronic correlations set the stage for the emergence of a variety of behaviors that are difficult to predict. The study of electronic behavior in Ce, Yb, U, and Pu intermetallic compounds with proximate magnetism has been particularly fruitful~\cite{Stewart:1984gj,Stewart:2001ks,Stewart:2006bz,Stockert:2012fo,Shirer:2012de,Bauer:2015jo}. In these materials, $f$ electrons may be incorporated into the Fermi surface as a result of hybridization with the conduction electrons. Consequently, these strongly correlated electron materials exhibit Kondo physics, magnetism, and unconventional superconductivity. In many cases, it has been shown that superconductivity emerges upon driving a particular strongly correlated system across the quantum critical point from the Kondo-dominated to the magnetic-dominated regime or vice versa with applied pressure, chemical substitution, or some other nonthermal tuning parameter~\cite{Si:2010ef,Gegenwart:2008de}. It is thought that magnetic fluctuations near this quantum critical point are key to generating an attractive interaction between electrons in these unconventional superconductors~\cite{White:2015bw,Monthoux:2007ha}.

The recently discovered family of heavy fermion materials Ce$_2M$Al$_7$Ge$_4$, where $M$ = Co, Ir, Ni, or Pd, crystallize in the noncentrosymmetric tetragonal space group $P\overline{4}2_1m$~\cite{Ghimire:2016fl}. These materials display heavy-electron behavior as well as a magnetic transition of unknown wave vector. The magnetic transition temperature is suppressed with chemical substitution such that $T_M$ = 1.8, 1.6, and 0.8 K, and no magnetic order above 0.4 K for $M$ = Co, Ir, Ni, and Pd respectively, indicating that the ``2174'' family of materials may be close to a magnetic quantum critical point~\cite{Ghimire:2016fl}. Many Ce-based materials have been shown to be close to a magnetic quantum critical point where unconventional superconductivity often emerges~\cite{Weng:2016if}.

Electron-nucleus interactions enable nuclear magnetic resonance (NMR) experiments to probe the ground state of the electronic system on a local level. NMR is sensitive to the local magnetic susceptibility/magnetic ordering via the hyperfine interaction and to the local electronic structure via the nuclear quadrupole interaction of nuclei that have spin $I > \frac{1}{2}$ with the electric field gradient (EFG) at the nuclear site. The resulting NMR Knight shift and quadrupolar splitting of the satellite transitions allow for the identification of nonequivalent atomic sites and provide microscopic information. NMR has been shown to be a powerful probe of magnetism, superconductivity, and the normal state of heavy fermion materials~\cite{Curro:2009fsa,Curro:2016cy}.

Here we present an NMR study of the spectral and dynamical properties of ${}^{59}$Co and ${}^{27}$Al in \ccag{}. We have identified an NMR Knight shift ($K$) anomaly indicating the onset of coherence of the Kondo lattice below a characteristic temperature $T^*$, below which the NMR Knight shift and the bulk magnetic susceptibility are no longer proportional~\cite{Curro:2016cy}. We also measured the spin-lattice relaxation rate (\slrr{}) as a function of temperature. We find that at high temperatures, $(T_1T)^{-1} \propto T^{-1}$ indicating that $\mathbf{q}=0$ spin fluctuations of the local moments dominate the relaxation. Estimates of the high-temperature limiting \slrr{} driven by local moment spin fluctuations agree well with the experimentally observed value. We also find that $(T_1TK)^{-1} \propto T^{-1/2}$ at the ${}^{59}$Co site as expected for a Kondo system at high temperatures relative to $T^*$ and $T_K$~\cite{Cox:1985cm}. At low temperatures, ${}^{59}$Co $T_1^{-1}$, measured by nuclear quadrupole resonance (NQR), diverges indicating slowing down above the magnetic ordering temperature $\theta = 1.65$ K. We find that the spin fluctuations are anisotropic at the ${}^{59}$Co site, with larger fluctuations in the basal plane that increase at low temperatures. Furthermore, our measurements of the bulk magnetic susceptibility versus temperature indicate weak ferromagnetic correlations above the magnetic transition.

\section{Experimental Details}
\label{sec:experimental_details}

Single crystalline samples of Ce$_2$CoAl$_7$Ge$_4$ were synthesized by the flux method as reported previously~\cite{Ghimire:2016fl}. A single crystal with the $c$ axis perpendicular to the largest facet of the plate-like morphology with dimensions $2.5 \times 2.5 \times 0.25$ mm$^3$ was selected for the NMR/NQR measurements. The sample was mounted on a microscope cover slip using GE7031 varnish and aligned based on the crystalline facets for NMR measurements. The orientation was confirmed based on the NMR spectra and theoretical calculations of the local electric field gradient tensor orientation at the $^{27}$Al ($I=5/2$, $^{27}\gamma/2 \pi = 11.0943$ MHz/T, $Q = 0.149$ b) and $^{59}$Co ($I=7/2$, $^{59}\gamma/2 \pi = 10.054$ MHz/T, $Q = 0.4$ b) atomic sites as discussed below~\cite{Walstedt:1967ca,Spiess:1969jk,Godbout:1997kz,Harris:2001dv,Harris:2008eh}.

NMR and NQR measurements were performed using commercial phase coherent spectrometers from Tecmag Inc. All spectral and $T_2^{-1}$ data were collected by performing an optimized $\pi/2 - \tau - \pi$ spin-echo pulse sequence. \slrr{} was measured by integration of the phase corrected real part of the spin echo following an inversion recovery $\pi - t_{wait} - \pi/2 - \tau - \pi$ pulse sequence. The resulting data were fit to the corresponding ($I=7/2$ central transition for ${}^{59}$Co, and $I=5/2$ central transition and first satellite for ${}^{27}$Al) multiexponential normal modes relaxation equation appropriate for a quadrupolar split spectrum subject to magnetic relaxation~\cite{Narath:1967do}. Home-built NMR probes with tunable cryogenic capacitors were used in conjunction with superconducting NMR magnet systems built by Cryomagnetics, which have a field homogeneity better than 10 ppm over a 1 cm diameter spherical volume. Field and frequency sweep modes of operation were employed for $H_0 \parallel \hat{a}$ and $H_0 \parallel \hat{c}$, respectively. For the frequency swept spectra, a home-built auto-tuning setup was employed to collect data, which allowed for tuning and matching of the NMR tank circuit over the entire frequency swept range.

Low-temperature magnetic susceptibility measurements were collected using a SQUID-based magnetometer with a ${}^3$He module with a base temperature of 420 mK.

\section{Calculation Details}
\label{sec:calculation_details}

Density functional theory (DFT) calculations~\cite{Hohenberg:1964fz} were performed using the WIEN2k code~\cite{Schwarz:2003cd}. The Perdew-Burke-Ernzerhof exchange-correlation potential based on the generalized gradient approximation~\cite{Perdew:1996iq} was used, and spin-orbit interactions were included through a second variational method. We performed calculations first by treating the $f$ electron as an itinerant valence electron, and subsequently by explicitly localizing the $f$ electron as a core state.

\section{NMR and NQR Spectra}
\label{sec:NMR_NQR_spectra}

\begin{figure}
	\includegraphics[width=\linewidth]{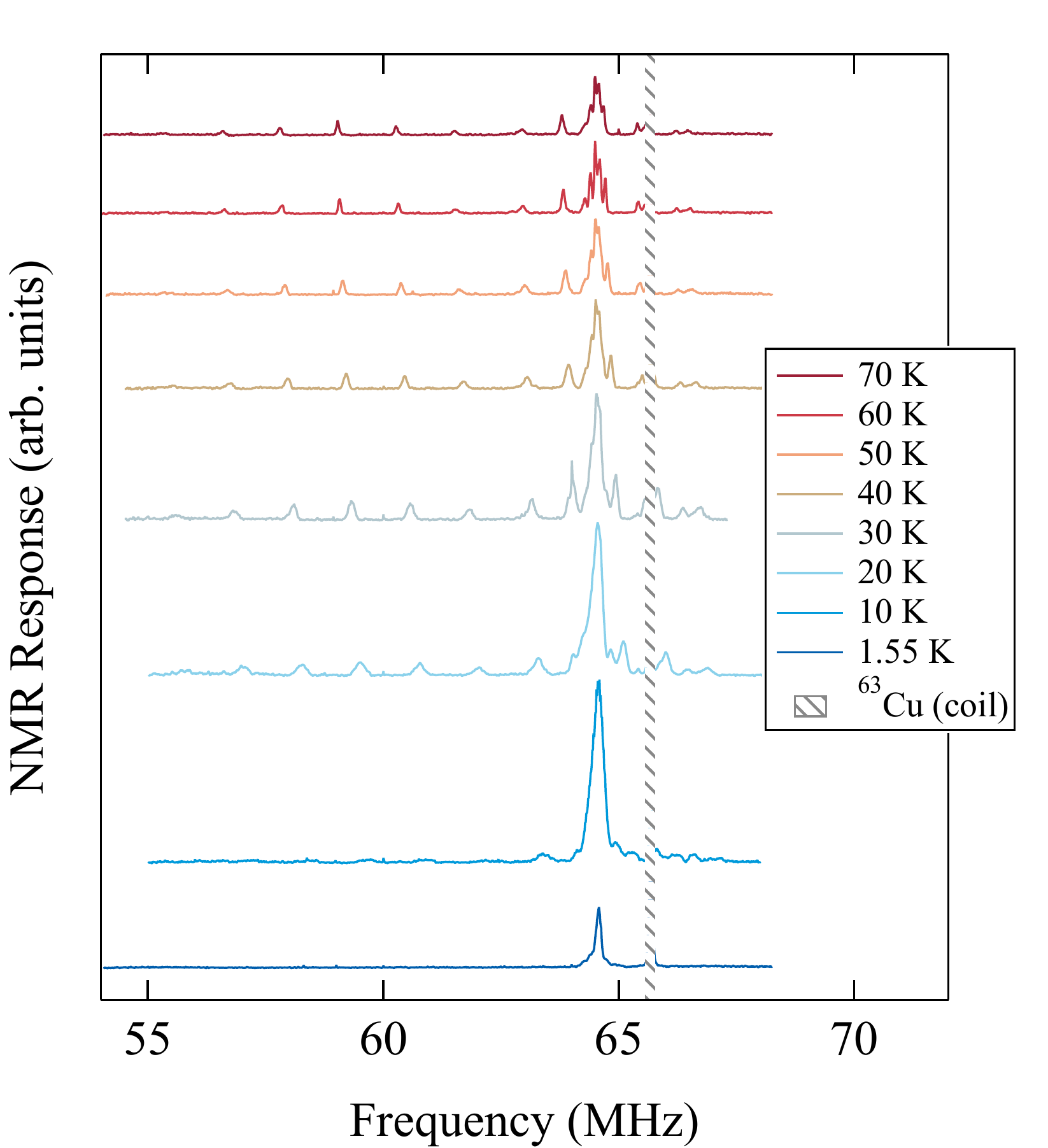}
	\caption{Frequency swept NMR spectra of \ccag{} at several temperatures for $H_0 \parallel \hat{c} = 5.8$ T.}
	\label{fig:FS_spectra}
\end{figure}

Frequency swept spectra of \ccag{} were acquired by performing a fast fourier transform (FFT) sum of evenly spaced frequency-swept spin echoes in an external field of 5.8~T for $H_0 \parallel \hat{c}$. The resulting temperature-dependent spectra are shown in Fig.~\ref{fig:FS_spectra}.
The seven lowest frequency resonances correspond to $^{59}$Co, and the higher frequency resonances are a superposition of $^{27}$Al(1-4). The $^{59}$Co site has tetragonal symmetry, with the principal axes of the EFG tensor lying along the crystalline axes, and with the largest component of the EFG along $\hat{c}$. The four $^{27}$Al sites have lower symmetries and EFG tensor orientations that do not coincide with the crystalline axes. A comparison of the experimental spectra with calculated EFGs from DFT will be discussed later.

\begin{figure}
	\includegraphics[width=\linewidth]{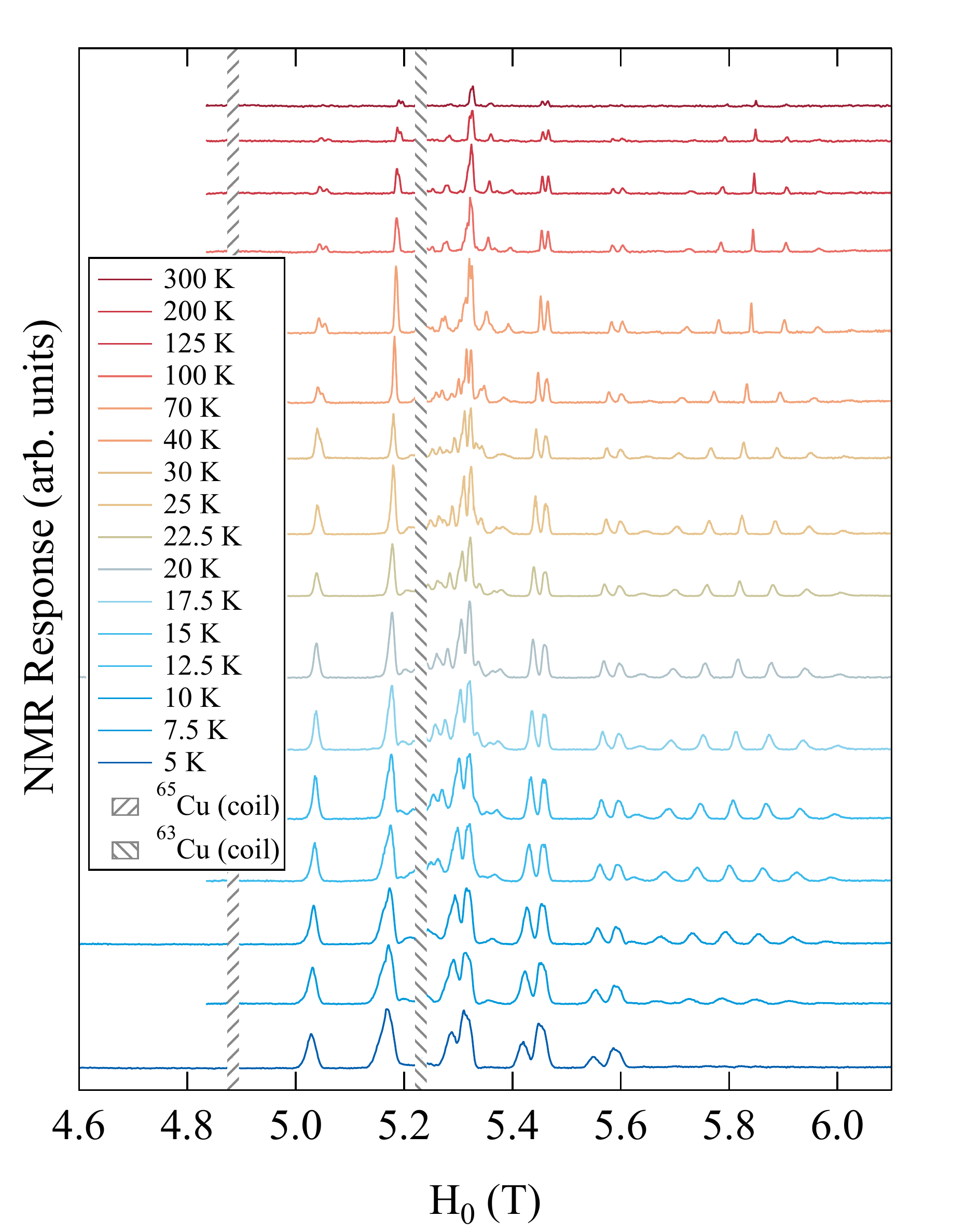}
	\caption{Field swept NMR spectra of \ccag{} at several temperatures for $H_0 \parallel \hat{a}$ at a fixed frequency $f_0 = 59.147$ MHz.}
	\label{fig:HS_spectra}
\end{figure}

Field swept spectra were collected for $H_0 \parallel \hat{a}$ at a fixed frequency of 59.147 MHz and are shown as a function of temperature in Fig.~\ref{fig:HS_spectra}. In this case, the field is oriented perpendicular to the largest component of the EFG tensor for $^{59}$Co (the seven evenly spaced resonances at higher field), but closer to the principal axes of the Al(3) and Al(4) sites. This conclusion is based on spectral simulations from exact diagonalization of the nuclear spin Hamiltonian as discussed below.

Based on the low natural abundance of ${}^{73}$Ge (7.7\%) and its low gyromagnetic ratio (${}^{73}\gamma/2\pi = 1.4852$ MHz/T), we do not expect to be able to observe the Ge nuclei in our field or frequency swept NMR experiments in \ccag{}. If the EFGs at the Ge sites are large enough, then, given the high spin $I=9/2$ and appreciable quadrupole moment ($Q = -0.22$ b), it is possible that we could observe ${}^{73}$Ge NQR. The intensity, however, would be approximately an order of magnitude smaller than the Co and Al resonances and therefore difficult to find.

Our goal in acquiring these temperature-dependent spectra is to extract the electric field gradient and Knight shift at the Co and Al sites. To extract these parameters, one can treat the nuclear spin Hamiltonian in either second-order perturbation theory---when, for example, the quadrupole interaction is much weaker than the Zeeman interaction---or by exact diagonalization. The nuclear spin Hamiltonian is given by $\mathcal{H} = \mathcal{H}_{\mathrm{ZH}} + \mathcal{H}_{Q}$. In this equation, the Zeeman Hamiltonian has been combined with the hyperfine Hamiltonian and can be expressed as
\begin{equation}
	\label{eqn:ZeemanHamiltonian}
	\mathcal{H}_{\mathrm{ZH}} = \gamma \hslash (1 + K) \mathbf{H}_{\mathbf{0}} \cdot \hat{\mathbf{I}},
\end{equation}
where $\gamma$ is the gyromagnetic ratio, $\hslash$ is the reduced Planck's constant, $K = K_0 + K_s$ is the shift (with the temperature independent orbital shift $K_0$ and the Knight shift $K_s$), $\mathbf{H}_{\mathbf{0}}$ is the external applied magnetic field, and $\hat{\mathbf{I}} = \hat{I}_x\hat{x} + \hat{I}_y\hat{y} + \hat{I}_z\hat{z}$. In general, $K$ will be a tensor with the same symmetry as the lattice site, but in this case we treat it as a free parameter for the two experimental orientations of the single crystal.

The second term in the nuclear spin Hamiltonian describes the nuclear quadrupole interaction with the EFG at the nuclear site, and can be expressed as
\begin{equation}
	\label{eqn:QuadrupoleHamiltonian}
	\mathcal{H}_{Q} = \frac{e Q V_{zz}}{4I(2I - 1)} \left[3 \hat{I}_z^2 - I^2 +
	\eta \left( \hat{I}_x^2 - \hat{I}_y^2  \right)\right],
\end{equation}
where $e$ is the elementary charge, $Q$ is the nuclear quadrupole moment, $V_{zz} = eq$ is the principal component of the EFG tensor (where $q$ is the field gradient),~\cite{Kaufmann:1979gn} $I_\beta$ with $\beta \in \set{x,y,z}$ are the nuclear spin operators, $I$ is the nuclear spin of the nucleus, and $\eta = (V_{xx} - V_{yy})/V_{zz}$ is the asymmetry parameter. We also express the NQR frequency as
\begin{equation}
	\label{eqn:nu_c}
	\nu_Q = \frac{3 e Q V_{zz}}{2I(2I-1)h} \sqrt{1 + \frac{\eta^2}{3}},
\end{equation}
where $h$ is Planck's constant.

\begin{figure}
	\includegraphics[width=\linewidth]{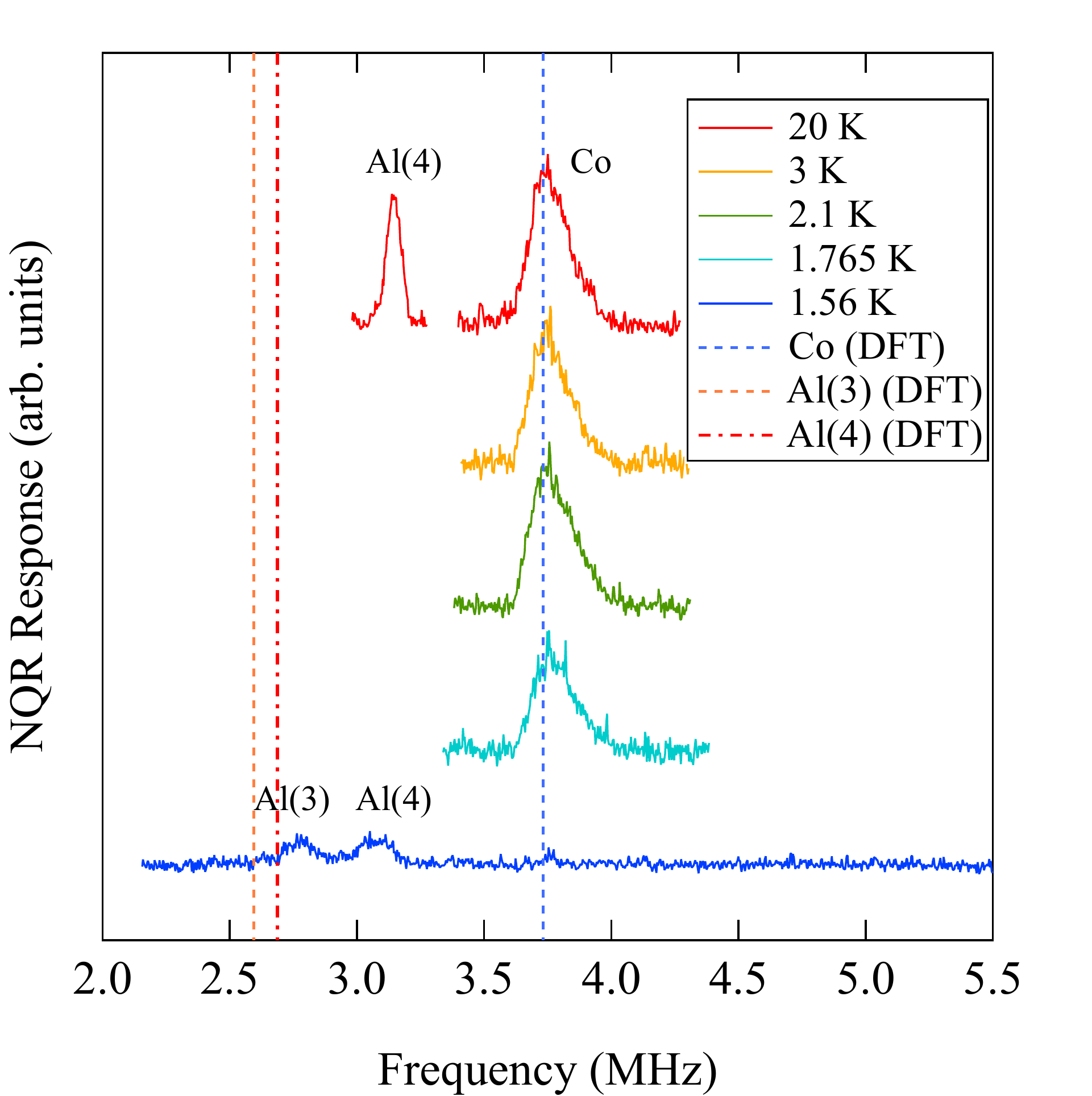}
	\caption{NQR spectra of \ccag{} at several temperatures. Dashed lines indicate the positions of the resonances as predicted by our localized $f$-electron DFT calculations.}
	\label{fig:NQR_spectra}
\end{figure}

NQR spectra of \ccag{} were collected at several temperatures and are shown in Fig.~\ref{fig:NQR_spectra}. Due to experimental difficulties of working in the low-frequency regime, we were only able to observe the highest frequency resonances of both the Co and Al NQR spectra. Due to fast relaxation at the Co site, we were barely able to discern the $3\nu_Q$ resonance of the Co NQR spectrum at the cryostat's base temperature of 1.56 K. The $2\nu_Q$ resonances of the Al(3) and Al(4) were still visible at 1.56 K, ostensibly due to the weaker hyperfine interaction at those sites as discussed below in Section~\ref{sec:Knight_shift_anomaly}.

\section{Comparison with DFT}
\label{sec:comparison_with_DFT}

\begin{figure} 
	\includegraphics[trim=0cm 0cm 12cm 0cm, clip=true, width=\linewidth]{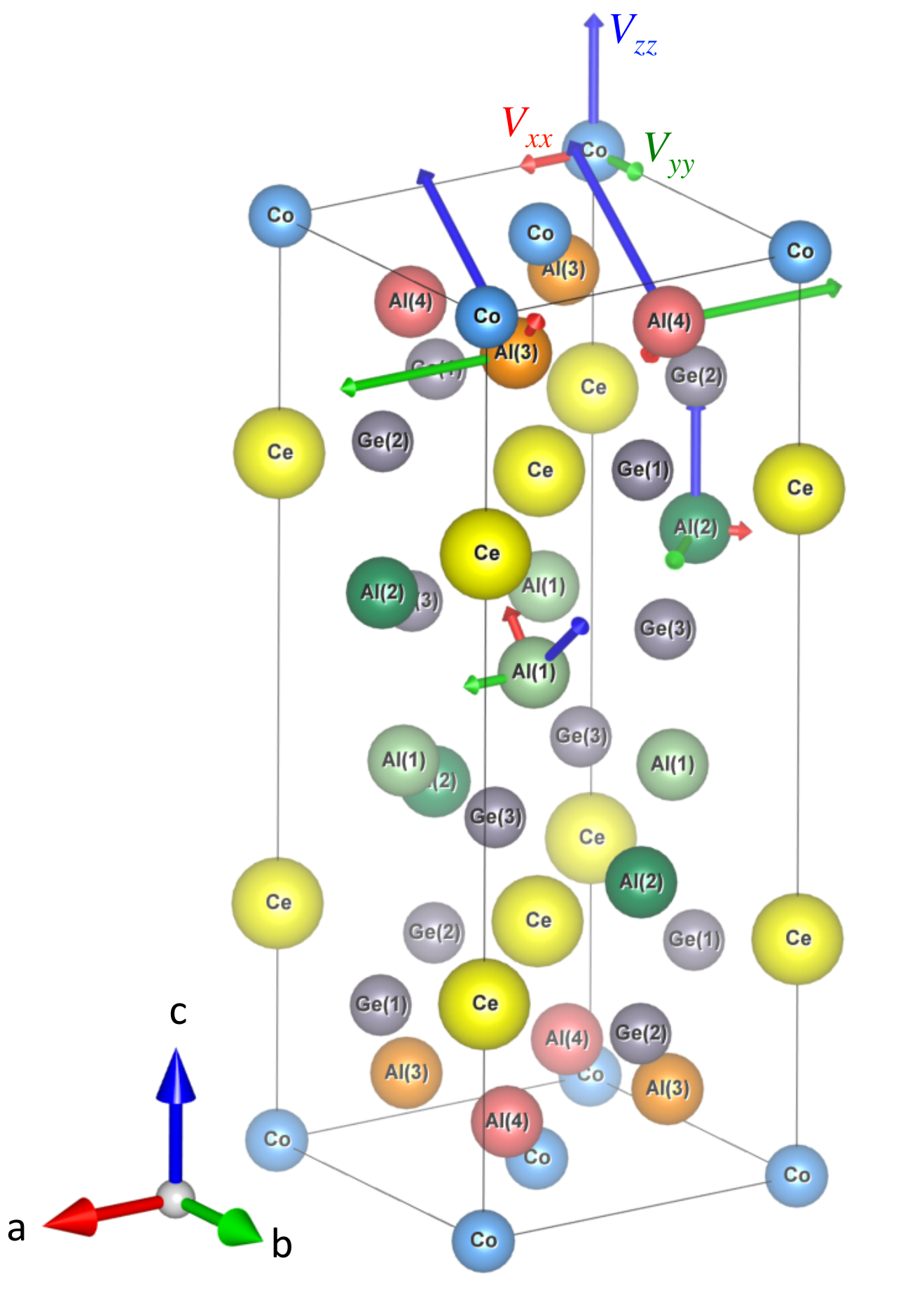}
	\caption{Crystal structure of \ccag{} with the principal axes of the EFG at the Co and Al(1-4) sites as calculated by DFT and scaled by the modulus of the tensor element. $V_{zz}$ (blue) is the largest component of the diagonal tensor, and the naming convention of $V_{xx}$ (red) and $V_{yy}$ (green) is such that $\eta = (V_{xx}-V_{yy})/V_{zz} > 0$~\cite{Momma:2011dd}.}
	\label{fig:EFG_dirs}
\end{figure}

We performed DFT calculations to extract the EFG at the nuclear sites using the WIEN2K software package as detailed in Section~\ref{sec:calculation_details}. The calculated EFG parameters for the relevant sites are shown in Table~\ref{tab:EFG_parameters}, and the corresponding EFG tensor orientations are shown in Fig.~\ref{fig:EFG_dirs}.
\begin{table}
\begin{ruledtabular}
\begin{tabular}{ccccc}
		& \multicolumn{2}{c}{$f$-electron localized}		& \multicolumn{2}{c}{$f$-electron itinerant}		\\
Site  	& $\lvert V_{zz} \rvert$ (MHz)	& $\eta$ (unitless)	& $\lvert V_{zz} \rvert$ (MHz) & $\eta$ (unitless)	\\
\hline
Co		& 1.252							& 0					& 1.755                        & 0 					\\
Al(1)	& 0.745							& 0.110				& 0.770                        & 0.012	 			\\
Al(2)	& 0.874							& 0.563				& 0.979                        & 0.422 				\\
Al(3)	& 1.628							& 0.828				& 1.728                        & 0.849				\\
Al(4)	& 1.643							& 0.777				& 1.736                        & 0.825 				\\
\end{tabular}
\end{ruledtabular}
\caption{\label{tab:EFG_parameters}DFT calculated EFGs at the observable sites. See Fig.~\ref{fig:EFG_dirs} for the spatial orientations of the EFG tensors.}
\end{table}

We find that allowing the Ce $f$ electron to become itinerant results in an overall enhancement of the EFG at all sites as shown in the rightmost columns of Table~\ref{tab:EFG_parameters}. At the Co site the EFG in the itinerant Ce $f$-electron calculation is 41\% larger than the observed value. The EFG at the Al sites remains reasonably close to the experimentally observed values, though slightly overestimated. Previous comparisons of calculations of the EFG at the In sites in CeIn$_3$~\cite{Lalic:2001jr,JalaliAsadabadi:2007jh,Kohori:1999kn} and Ce$M$In$_5$ ($M$ = Co, Rh, and Ir)~\cite{Rusz:2008im} were also able to provide reasonable agreement between DFT and experimental observations of the EFG by NQR. In the case of the 115 heavy fermion materials the DFT calculations resulted in better agreement when the $f$-electron was localized as opposed to participating in the Fermi surface. Indeed, in the case of \ccag{}, we find better agreement between experiment and theory at the Co site and similar agreement at the Al sites upon localizing the Ce $f$ electron. NMR measurements and bulk measurements also indicate local moment behavior~\cite{Ghimire:2016fl}. In the comparison to experiment below we will restrict ourselves to the calculations for which the $f$ electron is localized.

Comparison of the NQR spectra with the DFT simulated resonance frequencies of the $3\nu_Q$ resonance for Co and the $2\nu_Q$ resonances for Al(3) and Al(4) are shown in Fig.~\ref{fig:NQR_spectra}. The theoretical predictions for the frequencies of these resonances are shown as dashed vertical lines. The predicted EFG and therefore the $3\nu_Q$ transition for ${}^{59}$Co is within 1\% of the observed value. This represents a success of the DFT code, as the value of the EFG is often difficult to predict, especially when heavy elements such as Ce are involved. The predicted values of the EFGs for Al(3) and Al(4) are not as close to the observed values. For Al(4) at 20 K the predicted frequency of the $2\nu_Q$ transition is reduced by 15\% from the observed value. This deviation is reduced to 12.4\% at 1.56 K. The Al(3) $2\nu_Q$ transition was only measured at the lowest temperature T = 1.56 K due to experimental difficulties of working in the low-frequency regime. At this temperature the calculated resonance frequency was suppressed by 6.6\%.

\begin{figure} 
	\includegraphics[trim=0.4cm 0cm 1cm 0cm, clip=true, width=\linewidth]{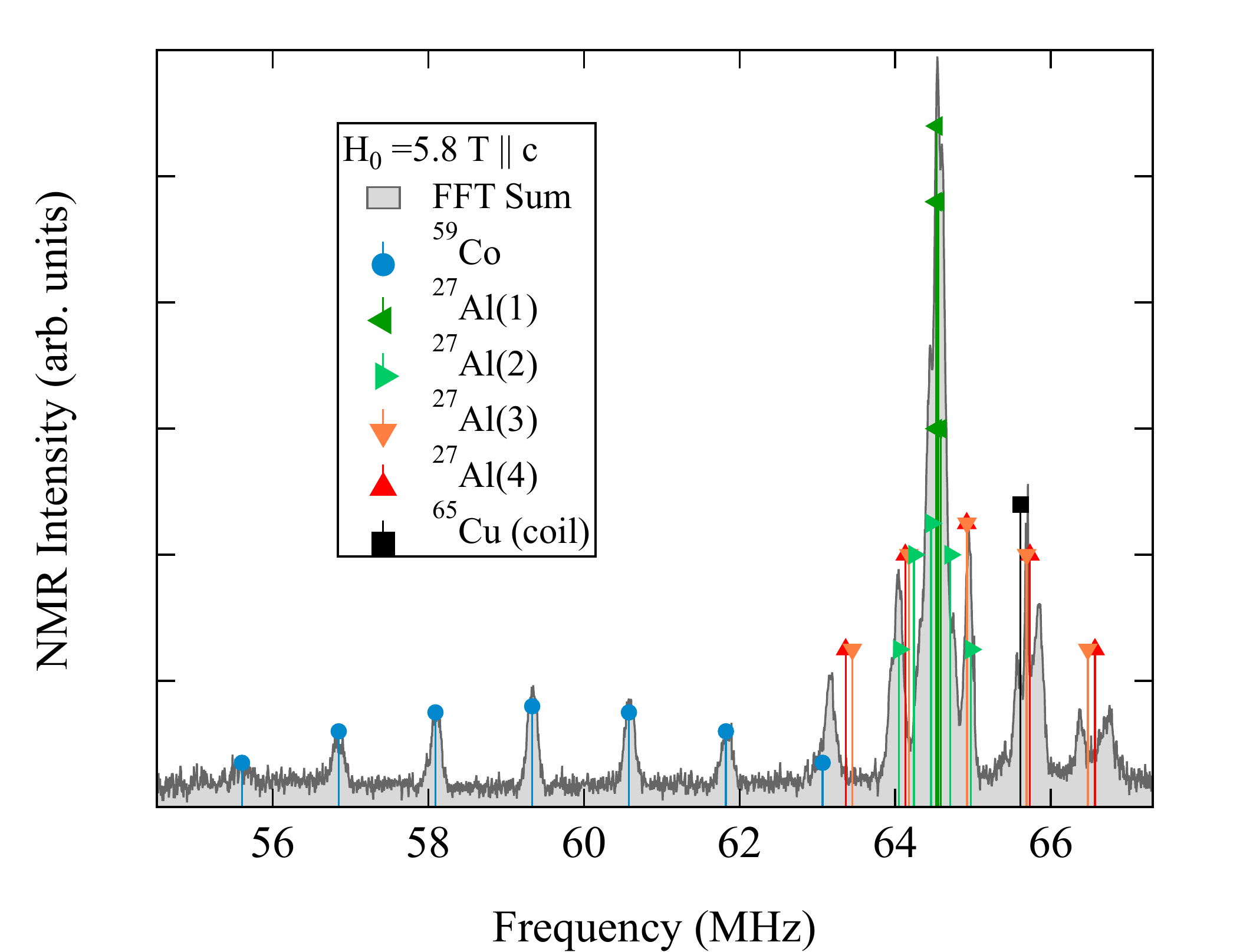}
	\includegraphics[trim=0cm 0cm 1cm 0cm, clip=true, width=\linewidth]{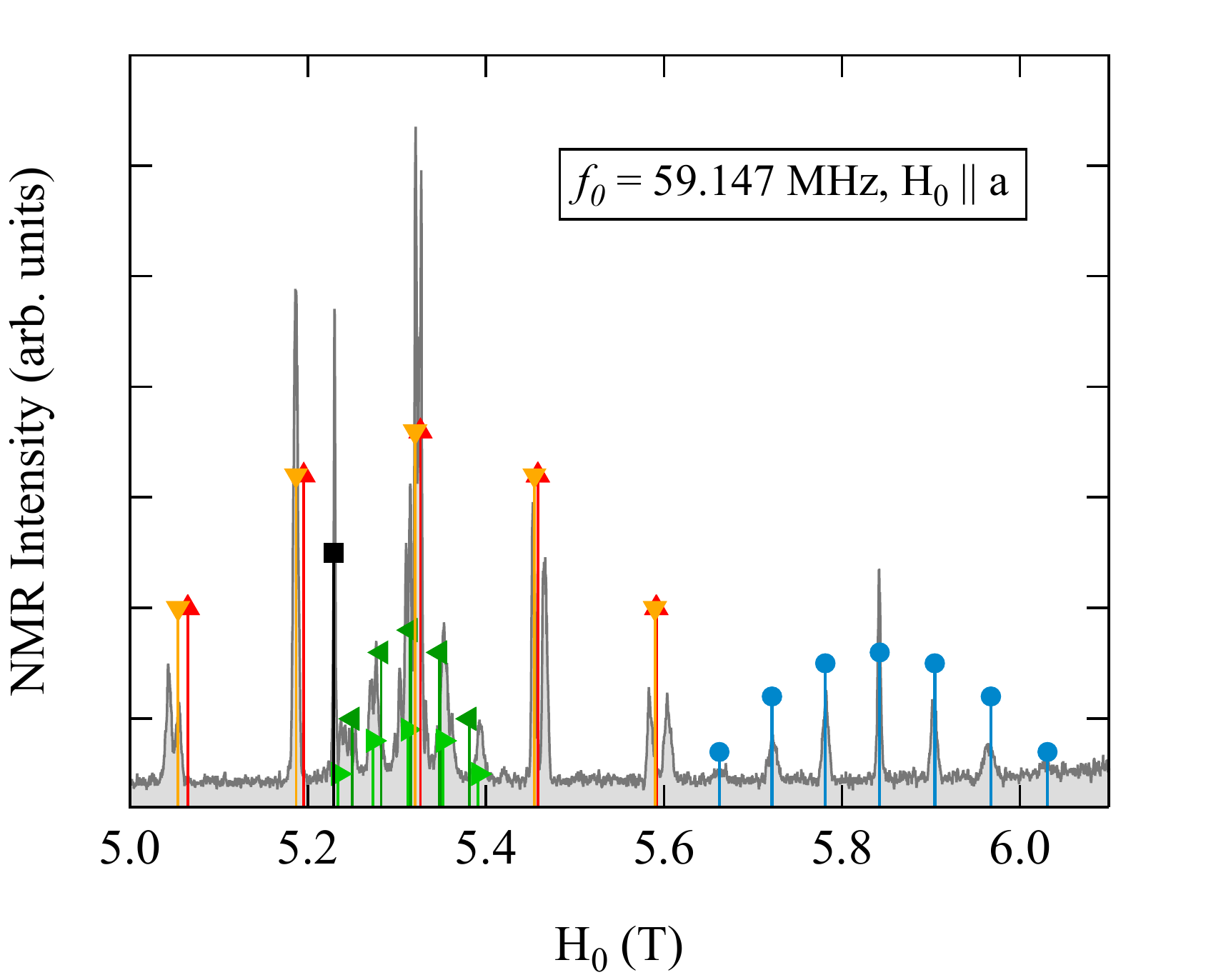}
	\caption{Comparison of experimental NMR spectra of \ccag{} with simulated spectral lines based on the EFGs calculated by DFT (localized $f$-electron case) for (top, frequency sweep) $H_0 \parallel \hat{c}$ at $T = 30$ K and (bottom, field sweep) for $H_0 \parallel \hat{a}$ at $T = 70$ K.}
	\label{fig:HpC_thry_exp}
\end{figure}

The lower symmetry of the Al sites results in a misalignment of the principal axes of the EFG tensors as shown in Fig.~\ref{fig:EFG_dirs}. We compare the theoretical calculations to the experimentally observed spectra in Fig.~\ref{fig:HpC_thry_exp}. The top panel shows a direct comparison of the calculated frequency swept spectrum at 30 K for $H_0 = 5.8\mathrm{~T} \parallel \hat{c}$, and bottom panel shows the field swept spectrum at 70 K for $H_0 \parallel \hat{a}$. We note that in the case of the Al sites, the experimental EFG values are somewhere in between the localized and itinerant calculations. Overall, the agreement between theory and experiment here is quite good, though a full characterization of the EFGs at the low-symmetry Al sites would require systematic rotation studies that are beyond the scope of this work.

\section{Knight Shift Anomaly}
\label{sec:Knight_shift_anomaly}

A key universal behavior that is observed in a variety of heavy fermion compounds is the Knight shift anomaly~\cite{Curro:2004bg}. In uncorrelated paramagnetic compounds, we expect the Knight shift to be proportional to the bulk magnetic susceptibility~\cite{Clogston:1961hh}. In Kondo lattice heavy fermion systems, there is a ubiquitous observation of deviation from this proportionality that has come to be associated with the formation of the coherent heavy electron fluid below the coherence temperature $T^*$~\cite{Curro:2016cy}. In the discussion below, we follow the literature and refer to the entire shift as the Knight shift. Technically, the term ``Knight shift'' is the shift from the expected resonant frequency of a nucleus in a metallic sample as compared to the same nucleus in a non-metallic sample due to the Pauli paramagnetism of the conduction electrons and the hyperfine coupling to the $s$-electron wave functions, which have appreciable overlap with the nucleus~\cite{Knight:1949ku,slichter1990principles}.

\begin{figure}
	\includegraphics[width=\linewidth]{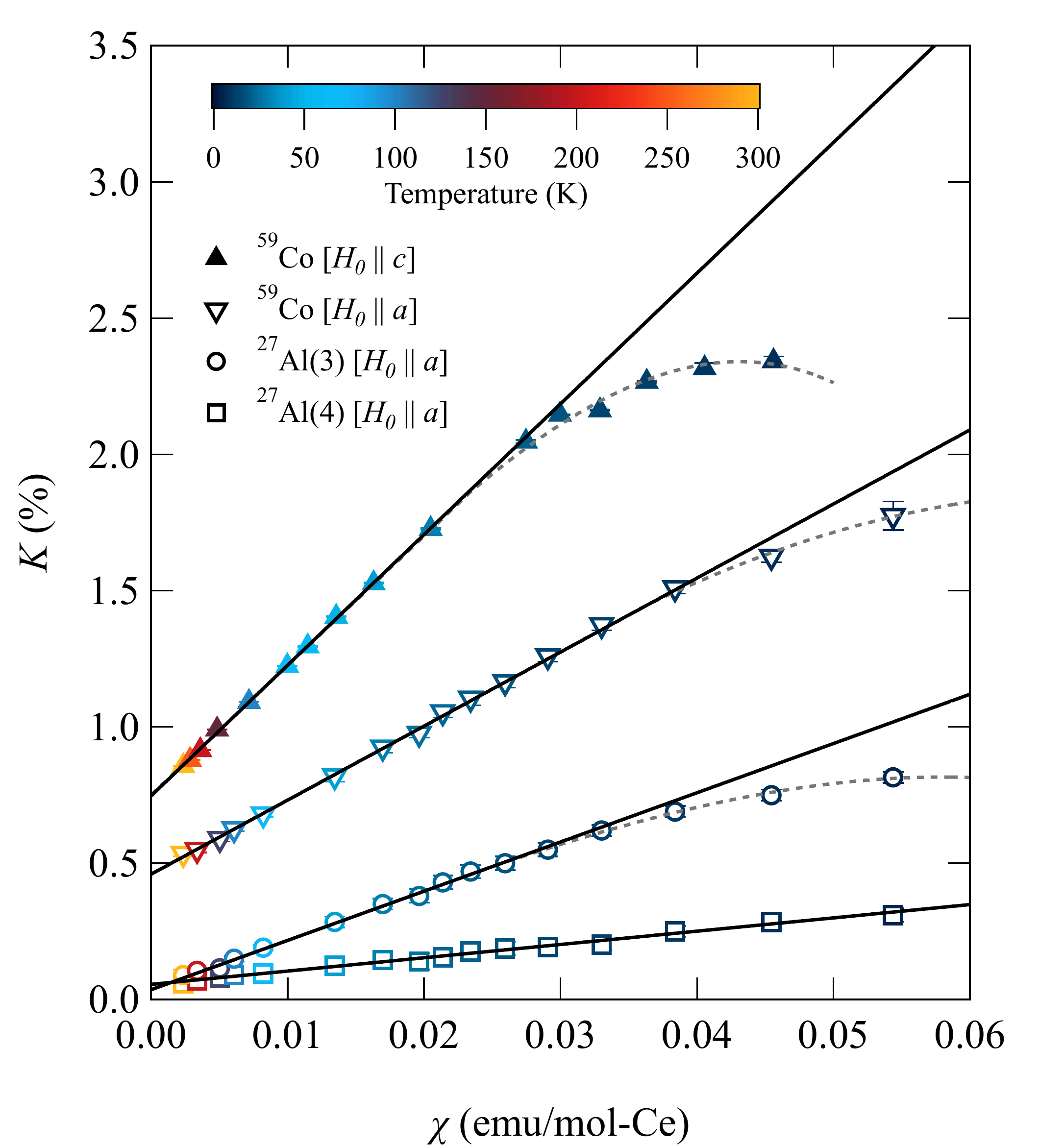}
	\caption{$K$ versus $\chi$ of \ccag{} with temperature as an implicit parameter for $H_0 \parallel \hat{c}$ and $H_0 \parallel \hat{a}$. Solid black lines are linear fits performed to extract the hyperfine couplings and temperature independent orbital shifts. Dashed grey curves are guides to the eye indicating the deviation of the Knight shift from the bulk susceptibility below the coherence temperature $T^*$.}
	\label{fig:K_vs_chi}
\end{figure}

We plot the Knight shift $K$ as a function of the magnetic susceptibility $\chi$ with temperature as an implicit parameter in Fig.~\ref{fig:K_vs_chi}. At high temperature, $K = A\chi + K_0$, where $A$ is the hyperfine coupling and $K_0$ is the temperature independent orbital shift. In general, the shift is a tensor quantity with contributions from both on-site and transferred hyperfine coupling contributions. In this case, due to the fact that the Ce $f$ electrons dominate the magnetic susceptibility, it is likely that the largest contribution is the transferred hyperfine interaction~\cite{Curro:2006dl}. We extract the high-temperature diagonal components of the Knight shift tensor $A_c$ and $A_a$ by performing linear least squares fits to the ${}^{59}$Co $K$ versus $\chi$ for both crystal orientations and ${}^{27}$Al(3) and ${}^{27}$Al(4) for $H \parallel \hat{a}$. The results of these fits are detailed in Table~\ref{tab:hyperfine_orbital}.

We find that the hyperfine coupling at the ${}^{59}$Co site is larger for $H \parallel \hat{c}$ than for $H \parallel \hat{a}$. Although the site symmetries and local environment of the ${}^{27}$Al(3) and ${}^{27}$Al(4) sites are nearly identical, these sites have different hyperfine couplings for $H \parallel \hat{a}$. This indicates that the hyperfine coupling at these sites is very sensitive to the local environment.

For $H \parallel \hat{c}$ we were not able to extract the hyperfine couplings at the ${}^{27}$Al(3) and ${}^{27}$Al(4) sites due to significant spectral overlap of the Al resonances in general. The ${}^{27}$Al(1) and ${}^{27}$Al(2) sites were also difficult to distinguish at enough temperature points for extraction of the hyperfine couplings for $H \parallel \hat{a}$. Hence, we focused our analysis of the Al spectra on the ${}^{27}$Al(3) and ${}^{27}$Al(4) sites.
\begin{table}
\begin{ruledtabular}
\begin{tabular}{cccc}
Site 			& Field Orientation 		& A (kOe/$\mu_B$)		& $K_0$ (\%) 			\\
\hline \\[-0.2cm]
${}^{59}$Co 	& $H_0 \parallel \hat{c}$ 	& 2.678 $\pm$ 0.017 	& 0.747 $\pm$ 0.003 	\\
${}^{59}$Co 	& $H_0 \parallel \hat{a}$ 	& 1.528 $\pm$ 0.024 	& 0.460 $\pm$ 0.005 	\\
${}^{27}$Al(3) 	& $H_0 \parallel \hat{a}$ 	& 1.009 $\pm$ 0.033 	& 0.036	$\pm$ 0.006		\\
${}^{27}$Al(4) 	& $H_0 \parallel \hat{a}$ 	& 0.273 $\pm$ 0.015 	& 0.056 $\pm$ 0.003 	\\
\end{tabular}
\end{ruledtabular}
\caption{\label{tab:hyperfine_orbital}Hyperfine coupling parameters $A$ and temperature independent orbital shifts $K_0$ for the measured field orientaitons and sites.}
\end{table}

At low temperatures, we find that the linear scaling breaks down at $T^* \sim 17.5 \pm 2.5$ K for $H \parallel \hat{c}$ and $T^* \sim 10 \pm 2.5$ K for $H \parallel \hat{a}$. Anisotropy in $T^*$ is quite common across a variety of heavy fermion compounds~\cite{Curro:2004bg}. For $H \parallel \hat{a}$, we find a Knight shift anomaly at the ${}^{27}$Al(3) site at $T^* \sim 12.5 \pm 2.5$. We find no Knight shift anomaly at the ${}^{27}$Al(4) site down to 5 K. This may be a consequence of the factor of three smaller hyperfine coupling at the ${}^{27}$Al(4) as compared to ${}^{27}$Al(3). It has been shown that the Knight shift anomaly nearly vanishes for $H \parallel \hat{a}$ in the Ce-115 systems~\cite{Curro:2016cy}. So, it is probable that the anomaly is weak at the ${}^{27}$Al(4) site for $H \parallel \hat{c}$ and then vanishes for $H \parallel \hat{a}$.

There are three possible mechanisms that are considered in the literature that account for the observed Knight shift anomaly in heavy fermion systems. The first possible explanation for the Knight shift anomaly is due to crystalline electric field (CEF) state depopulation below a characteristic temperature $T_{CEF}$~\cite{MacLaughlin:1981hh,Ohama:1995bh,Curro:2001kc,Ohishi:2009eq}. The second possibility is that the coherence of the Kondo lattice causes the local susceptibility at the nuclear site to differ from the bulk susceptibility below the coherence temperature $T^*$~\cite{Kim:1995eb,Yang:2008iw}. Indeed, both of these mechanisms have been shown to manifest in the Ce$_2$IrIn$_8$~\cite{Ohishi:2009eq}. The scale of $T_{CEF}$ is set by the energy splitting of the CEF states, which we expect to be similar to the CEF states of Ce$_2$PdAl$_7$Ge$_4$. The splitting between the lowest state and first excited CEF states in Ce$_2$PdAl$_7$Ge$_4$ is 100 K and is therefore a factor of 5--10 times larger than the observed values of $T^* \sim 10-17.5$ K. Furthermore, we note that the temperature range over which the magnetic entropy is released is on the order of $0 \leq T \leq T^*$~\cite{Ghimire:2016fl,Ohishi:2009eq}. Therefore, we associate the observed Knight shift anomalies with the development of coherence.

The third interpretation relies on treating the hyperfine coupling constant as temperature dependent. This picture was initially invoked in terms of discussing the Knight shift anomaly in conjunction with the dynamical susceptibility in CeIrIn$_5$~\cite{Kambe:2010dt} and CeCoIn$_5$~\cite{Sakai:2010ks}. However, it has also been argued that the energy scale of the mechanism (in this case orbital overlap/hopping integrals between the Ce and Co/Al wave functions) that generates the hyperfine coupling is much larger than those of the Kondo and CEF scales, and therefore would not be modified appreciably by these interactions~\cite{Abragam:1961vg,Shirer:2012de,Mila:1989gg}.

\section{Spin Dynamics}
\label{sec:spin_dynamics}

There are a variety of interactions that can bring the spin temperature of the nuclear ensemble into equilibrium with the lattice. In an uncorrelated metallic system the relaxation is dominated by spin-flip scattering with the conduction electrons. In a system with localized $f$ electrons, if the transferred hyperfine coupling to the nucleus under study is large, then fluctuations of the transferred hyperfine field will dominate the relaxation. The spin-lattice relaxation rate is also sensitive to magnetic fluctuations as they slow down and spectral weight becomes appreciable at the NMR/NQR frequency above a magnetic phase transition.

\begin{figure}
	\includegraphics[width=\linewidth]{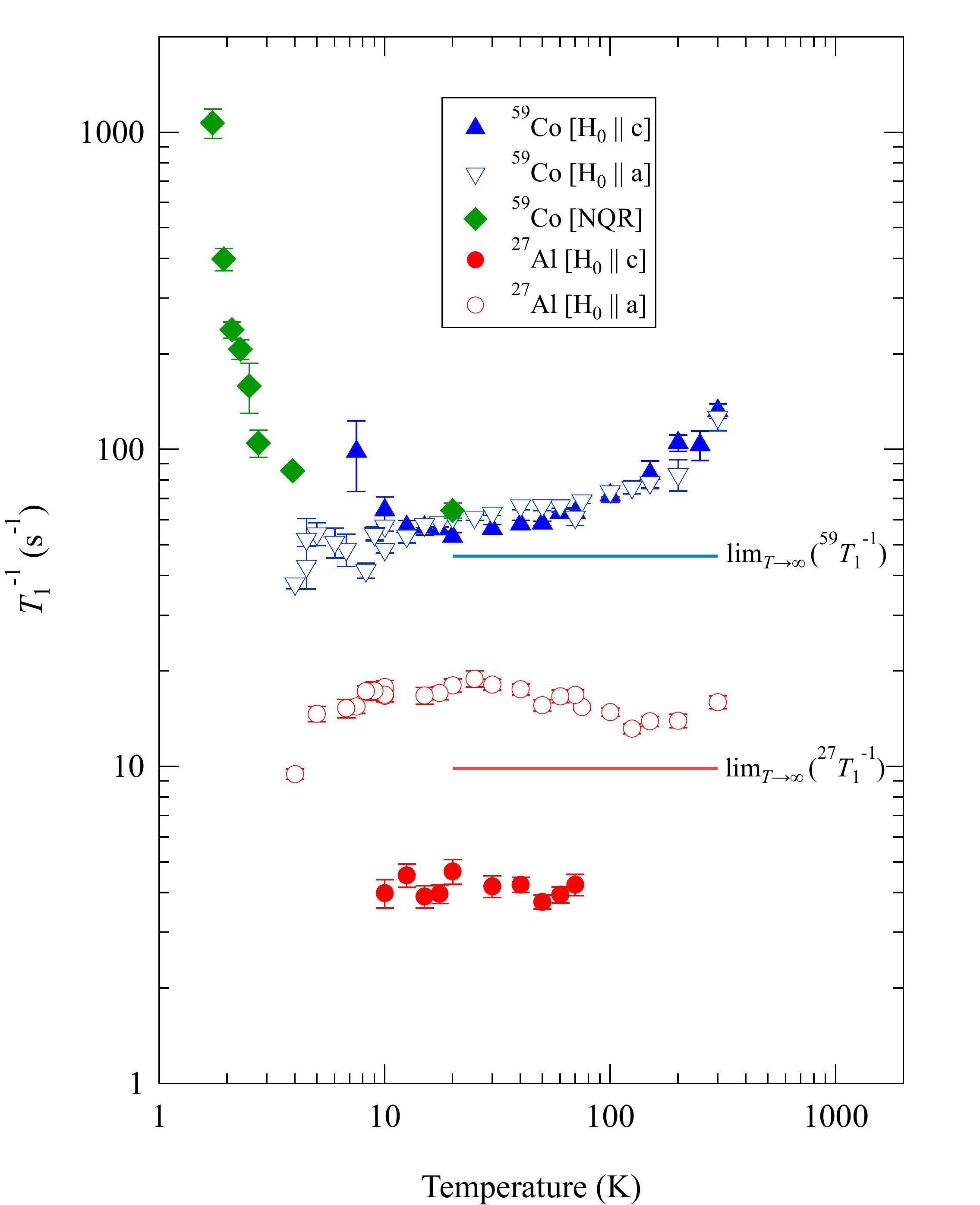}
	\caption{Spin-lattice relaxation rate $T_1^{-1}$ of \ccag{} versus temperature for $^{59}$Co and $^{27}$Al for $H_0 \parallel \hat{c}$ (filled triangles and circles respectively), $H_0 \parallel \hat{a}$ (open triangles and circles respectively), and zero field (NQR) $^{59}$Co (filled diamonds). Solid horizontal lines indicate the high-temperature limiting value of \slrr{} driven by local moment fluctuations as discussed in the text.}
	\label{fig:T1inv_vs_T}
\end{figure}

We measured the spin-lattice relaxation rate \slrr{} as a function of temperature at the ${}^{59}$Co site at the central transition for both $H_0 \parallel \hat{a}$ and $H_0 \parallel \hat{c}$. We also measured \slrr{} via ${}^{59}$Co NQR at the $3\nu_Q$ transition at low temperatures. For $H_0 \parallel \hat{a}$, we measured \slrr{} at the overlapping central transitions of Al(3) and Al(4), which include some contribution of the spectra from Al(1) and Al(2). However, we also measured \slrr{} at the first satellite transition on the high-field side of the spectrum of Al(3) and Al(4), and found that the value was equivalent to within the standard error. For the $H_0 \parallel \hat{c}$ \slrr{} measurements, there is a significant contribution of Al(1), and to a lesser degree Al(2), to the spectral weight near the central transitions of Al(3) and Al(4). Indeed, the inversion recovery curves exhibit stretched exponential behavior indicative of multiple relaxation rates. As such, we do not draw conclusions from the absolute magnitude or apparent anisotropy of the ${}^{27}$Al \slrr{} in general. The results of these measurements are summarized as the spin-lattice relaxation rate divided by temperature $(T_1T)^{-1}$ versus temperature in Fig.~\ref{fig:T1inv_vs_T}.

\begin{figure}
	\includegraphics[width=\linewidth]{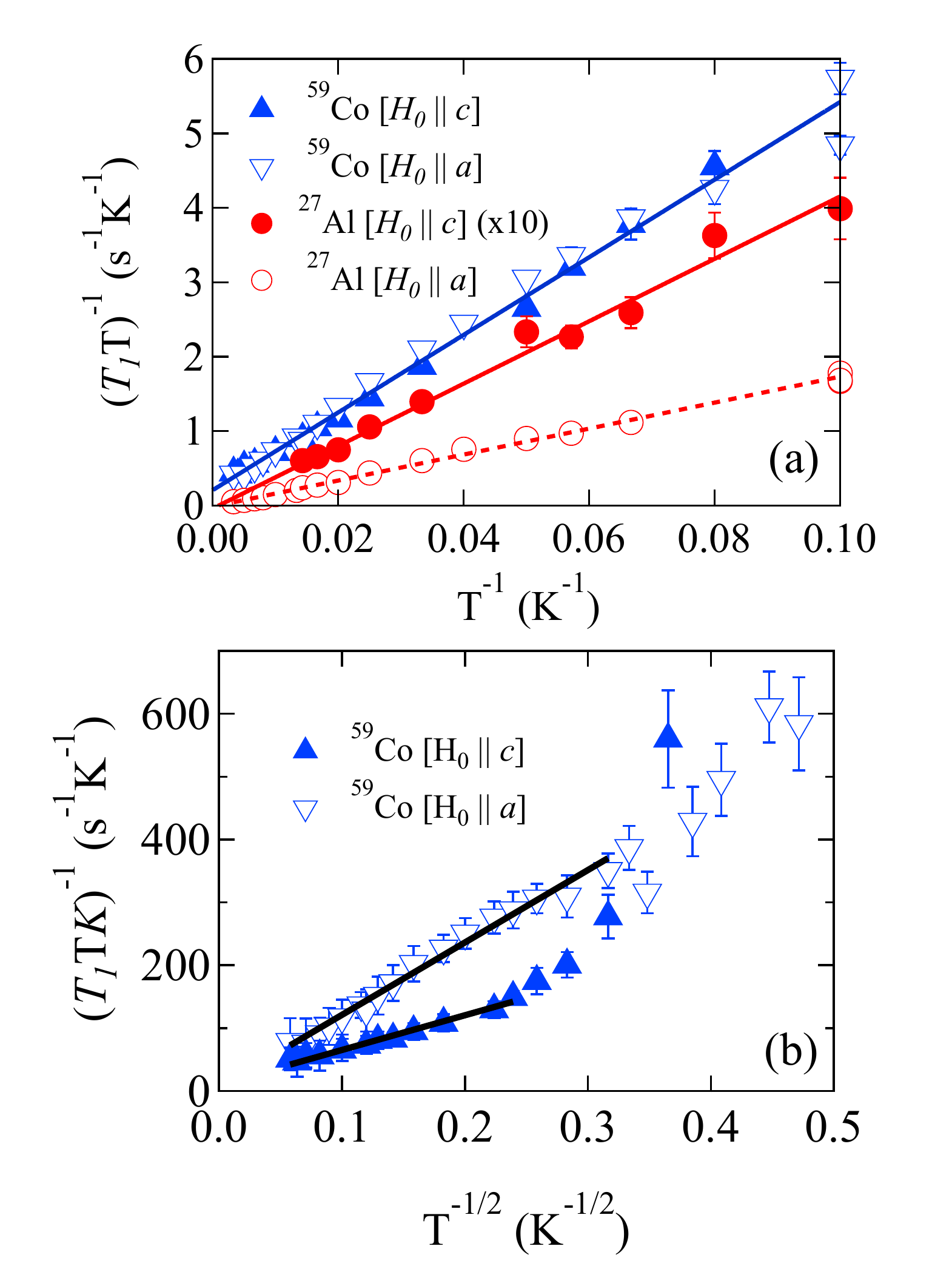}
	\caption{(a) Spin-lattice relaxation rate divided by temperature $(T_1T)^{-1}$ versus inverse temperature with linear fits indicating Curie-Weiss behavior in the high-temperature regime. (b) $(T_1TK)^{-1}$ versus $T^{-1/2}$, with linear fits indicative of Kondo behavior at high temperatures $T > T^*$).}
	\label{fig:T1Tinv_high}
\end{figure}

We interpret the data in two temperature regimes $T \gtrsim 20$ K and $T \lesssim 3$ K. At high temperatures, we find that $(T_1T)^{-1} \sim T^{-1}$ indicating Curie-Weiss behavior as shown in Fig.~\ref{fig:T1Tinv_high}(a). This scaling can be interpreted in terms of fluctuations of the local Ce moments dominating the relaxation with a weak metallic background. This behavior is similar to other Ce compounds at temperatures much higher than the exchange coupling between the moments (e.g., CePt$_2$In$_7$~\cite{Kohori:1999kn,Sakai:2011ez,Sakai:2014fp}). The metallic (Korringa) term in the relaxation is linear in temperature ($T_1^{-1} = a T$), and we find that it is quite small in comparison to the local moment driven relaxation over the temperature range measured.

The near temperature independence of \slrr{} at high temperatures relative to $T_M \sim 1.8 $K and $T_K \sim 5$ K indicates that the relaxation is dominated by fluctuations of the Ce $f$-electron local moments. We estimate the high temperature limit of the spin-lattice relaxation rate
\begin{equation}
\label{eqn:T1inv_LM_Moriya}
\lim_{T \rightarrow \infty}\left(\frac{1}{T_1}\right) = \sqrt{2\pi} \left(\frac{\gamma A g_J \mu_B}{z^\prime}\right)^2 \frac{z^\prime J(J+1)}{3 \omega_{ex}},
\end{equation}
from the theory of nuclear magnetic spin-lattice relaxation rates in a local moment antiferromaget~\cite{Moriya:1956hx}. In the above expression, $\gamma$ is the gyromagnetic ratio of the nucleus under study, $A=\sqrt{(2 A_a^2 + A_c^2)/3}$ is the isotropic hyperfine coupling constant, $g_J$ is the Land\'{e} $g$-factor, $\mu_B$ is the Bohr magneton, $z^\prime$ is the number of coupled ligand sites (Co or Al in this case), $J$ is the total angular momentum of the magnetic ion, and the exchange frequency
\begin{equation}
\label{eqn:omega_ex}
\omega_{ex} = \frac{k_B \Theta_{CW}}{\hbar} \sqrt{\frac{3 g_J}{z |g_J-1|^3 J(J + 1)}}.
\end{equation}
Here, $k_B$ is the Boltzmann constant, $\Theta_{CW} = -37$ K is the Weiss temperature from fits to bulk magnetic susceptibility measurements~\cite{Ghimire:2016fl}, and $z = 4$ is the number of nearest-neighbor local moments. In the current case of \ccag{}, we find that the bulk data~\cite{Ghimire:2016fl} are consistent with trivalent Ce$^{3+}$. As such, we use the appropriate values $J = 5/2$ and $g_J = 6/7$ in Eqn.~\ref{eqn:omega_ex} to find $\omega_{ex} = 9.19 \times 10^{12}$ s$^{-1}$.

Utilizing this calculated value of $\omega_{ex}$, ${}^{59}\gamma = 2 \pi 10.054$ MHz/T, $z^\prime = 2$ and ${}^{59}A=\sqrt{(2 ({}^{59}A_a)^2 + ({}^{59}A_c)^2)/3} = 0.199$ T/$\mu_B$ in Eqn.~\ref{eqn:T1inv_LM_Moriya} we estimate $\lim_{T \rightarrow \infty}\left(\frac{1}{{}^{59}T_1}\right) = 46.03$ s$^{-1}$. This value agrees quite well with the measured high temperature value of \slrr{} as shown by the solid blue line in in Fig.~\ref{fig:T1inv_vs_T}. This analysis has been applied to several magnetic $f$-electron systems to differentiate between dominantly local moment or itinerant electron magnetic fluctuations~\cite{Sakai:2007hl,Chudo:2013cj}. That is, if $T_1^{-1} \sim \lim_{T \rightarrow \infty}\left(\frac{1}{{}^{59}T_1}\right)$, then one can attribute the dominant relaxation mechanism to local moment spin fluctuations.  Whereas, if $T_1^{-1} < \lim_{T \rightarrow \infty}\left(\frac{1}{{}^{59}T_1}\right)$, then one can argue that the system is in the itinerant weak magnetic limit. This framework developed by Moriya is similar to Anderson's model of relaxation/spectral narrowing due fluctuations of exchange-coupled local moments~\cite{WAnderson:1954eh}, which has been used before in various contexts to explain the relaxation in local moment systems~\cite{Yasuoka:1983jy,Hiraoka:1992ff,Matsumura:2009eu}.

In the case of the local moment relaxation at the ${}^{27}$Al sites, this analysis is not so straightforward given the apparent anisotropy observed in the \slrr{} and the overlap of the central transitions as discussed above. In spite of this, we acquire an order-of-magnitude estimate for the Al high-temperature local moment relaxation by first taking the average of the measured hyperfine couplings ${}^{27}A_a$ at the Al(3) and Al(4) sites. We then assume that the anisotropy will be similar to the Co site to calculate ${}^{27}A_c$. We then calculate ${}^{59}A=\sqrt{(2 ({}^{59}A_a)^2 + ({}^{59}A_c)^2)/3} = 0.0833$ T/$\mu_B$. Using this value and ${}^{27}\gamma = 2 \pi 11.0943$ MHz/T we find $\lim_{T \rightarrow \infty}\left(\frac{1}{{}^{27}T_1}\right) = 9.86$ s$^{-1}$, which is shown as a solid red line in Fig.~\ref{fig:T1inv_vs_T}. This value is also quite close to the observed values even with our multiple assumptions. Overall, our \slrr{} data agree well with these estimates, which can be interpreted as strong evidence of the high-temperature local moment driven \slrr{}.

We can go further with the relaxation analysis assuming a dense Kondo lattice of Ce local moments. The dominant local moment spin-lattice relaxation can be derived from~\cite{Moriya:1956hx,Moriya:1963hp,Moriya:1974fm}
\begin{equation}
	\label{eqn:T1inv_dynamic_chi}
	\frac{1}{T_1} = \gamma^2 k_B T \lim_{\omega \rightarrow 0} \sum_{\mathbf{q}} A^2(\mathbf{q}) \frac{\chi''(\mathbf{q},\omega)}{\omega},
\end{equation}
where $\gamma$ is the nuclear gyromagnetic ratio, $k_B$ is the Boltzmann constant, $T$ is temperature, $\omega$ is frequency, $A$ is the $\mathbf{q}$-dependent hyperfine coupling, $\chi''(\mathbf{q},\omega)$ is the imaginary part of dynamical susceptibility at wave vector $\mathbf{q}$ and frequency $\omega$. The susceptibility is normalized to units of $N_Ag^2\mu_B^2$, where $N_A$ is the number of local moments in one mole (Avogadro's number), $g$ is the electron $g$ factor, and $\mu_B$ is the Bohr magneton~\cite{Moriya:1974fm}. Taking the limit and performing the sum~\cite{Cox:1985cm} results in
\begin{equation}
	\label{eqn:electron_flucts_chi}
	\lim_{\omega \rightarrow 0} \sum_{\mathbf{q}} A^2(\mathbf{q}) \frac{\chi''(\mathbf{q},\omega)}{\omega} = \frac{\chi_L(T)}{\Gamma(T)},
\end{equation}
where $\chi_L(T)$ is the the local static susceptibility and $\Gamma(T)$ is the spin fluctuation rate of the local moments. Substituting this expression into Eqn.~\ref{eqn:T1inv_dynamic_chi} and noting that the local static susceptibility is proportional to the Knight shift for $T > T^*$, we arrive at the expression
\begin{equation}
	\label{eqn:electron_flucts_K}
	\frac{1}{T_1} = \gamma^2 A^3 k_B T \frac{K(T)}{\Gamma(T)}.
\end{equation}
Cox \textit{et al.} find that $\Gamma(T) \propto \sqrt{T}$~\cite{Cox:1985cm}. Rearranging, dividing both sides by $T$, and dropping the constants in Eqn.~\ref{eqn:electron_flucts_K} results in
\begin{equation}
	\label{eqn:T1TKinv_vs_Ttom0p5}
	\frac{1}{T_1TK} \propto \frac{1}{T^{1/2}}.
\end{equation}

At the ${}^{59}$Co site, where we unambiguously extract $T_1^{-1}$ without contamination from other sites, we perform a direct comparison to the expected theory for relaxation due to $f$-electron dominated relaxation in a Kondo lattice. For temperatures $T > T_K \sim 5$ K~\cite{Ghimire:2016fl} and $T > T^* \sim 20$ K where $K \propto \chi$, we expect Eqn.~\ref{eqn:T1TKinv_vs_Ttom0p5} to hold. Fig.~\ref{fig:T1Tinv_high}(b) shows $(T_1TK)^{-1}$ versus $T^{-1/2}$ with the linear fits indicating good agreement with behavior expected for a Kondo system above $T^*$ for both crystalline orientations. Similar behavior has been observed in several other Kondo systems at high temperatures in the local moment regime~\cite{MacLaughlin:1989dv,Bruning:2008kx,Aarts:1983df,Mahajan:1998bx}.

\begin{figure} 
	\includegraphics[trim=0cm 0cm 1cm 0cm, clip=true, width=\linewidth]{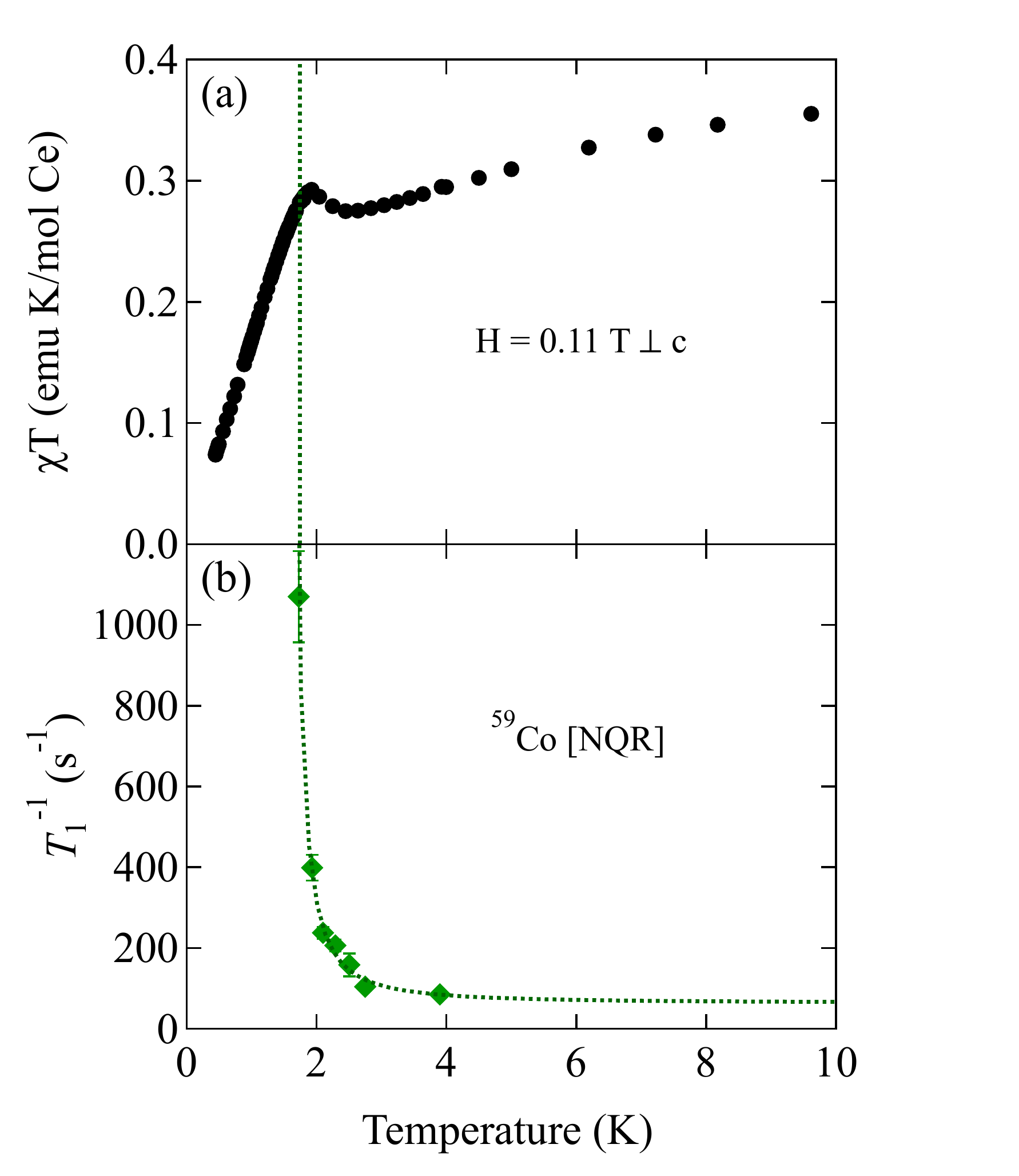}
	\caption{(a) Magnetic susceptibility of \ccag{} times temperature $\chi T$ versus temperature. (b) Spin-lattice relaxation rate $(T_1)^{-1}$ versus temperature measured by Co NQR. The dotted line is a guide to the eye indicating diverging spin fluctuations as discussed in the text.}
	\label{fig:chiT_T1inv_NQR}
\end{figure}

In the low-temperature regime ($T \lesssim 3$ K), we find evidence for slowing down of spin fluctuations as evidenced by a divergence of \slrr{} above the magnetic transition as shown in Fig.~\ref{fig:chiT_T1inv_NQR}(b). Based on the orientation of the EFG tensor at the Co site, relaxation as measured by NQR will probe fluctuations in the basal plane. We do not attempt to analyze this slowing down in a quantitative way as there are too many unknowns to utilize this evidence alone for dimensionality or character of the spin fluctuations above the magnetic transition. However, we show a dotted line guide to the eye in Fig.~\ref{fig:chiT_T1inv_NQR} extending in part (a), which agrees well with the phase transition at $T_M$ as observed in $\chi T$ versus temperature.

The upturn in $\chi T$ starting just above 2 K is indicative of weak ferromagnetic-like behavior where $\chi$ increases faster than 1/T with decreasing temperature just above the magnetic transition due to fluctuations near $\mathbf{q}=0$ (e.g., UGe$_2$~\cite{Onuki:1992cc}). Bulk magnetic susceptibility measurements find a significantly smaller effective moment in \ccag{} and no hysteresis as compared to Ce$_2$NiAl$_7$Ge$_4$ and Ce$_2$IrAl$_7$Ge$_4$, which both display ordering that is most likely ferromagnetic~\cite{Ghimire:2014dd}. Taken together, magnetic susceptibility and slow spin dynamics as observed via \slrr{} indicate that the fluctuations are of a weakly ferromagnetic character at low temperatures above the magnetic phase transition.

The lack of anisotropy in $T_1^{-1}$ at the Co site combined with the measured anisotropic hyperfine coupling indicates that the spin fluctuations are actually anisotropic. Spin-lattice relaxation is driven by fluctuations of the components of the hyperfine field perpendicular to the applied magnetic field. One can decompose the spin lattice relaxation rate for a given field alignment as follows:
\begin{equation}
	\label{eqn:T1Tinv_HpC}
	{(T_1T)^{-1}}_{H \parallel c} = 2 R_a
\end{equation}
and
\begin{equation}
	\label{eqn:T1Tinv_Hpa}
	{(T_1T)^{-1}}_{H \parallel a} = R_a + R_c,
\end{equation}
where, assuming the hyperfine form factors are unity, 
\begin{equation}
	\label{eqn:Ri}
	R_i \propto A_i^2 \lim_{\omega \rightarrow 0} \sum_\mathbf{q}\frac{\chi''_i(\mathbf{q},\omega)}{\omega},
\end{equation}
where $i = a, c$~\cite{Sakai:2017fj}. Using the above equations, we can therefore extract a  hyperfine-coupling-normalized quantity $R_i/A_i^2$, which is proportional to the imaginary part of the dynamical susceptibility summed over the first Brillouin zone.

\begin{figure}
	\includegraphics[width=\linewidth]{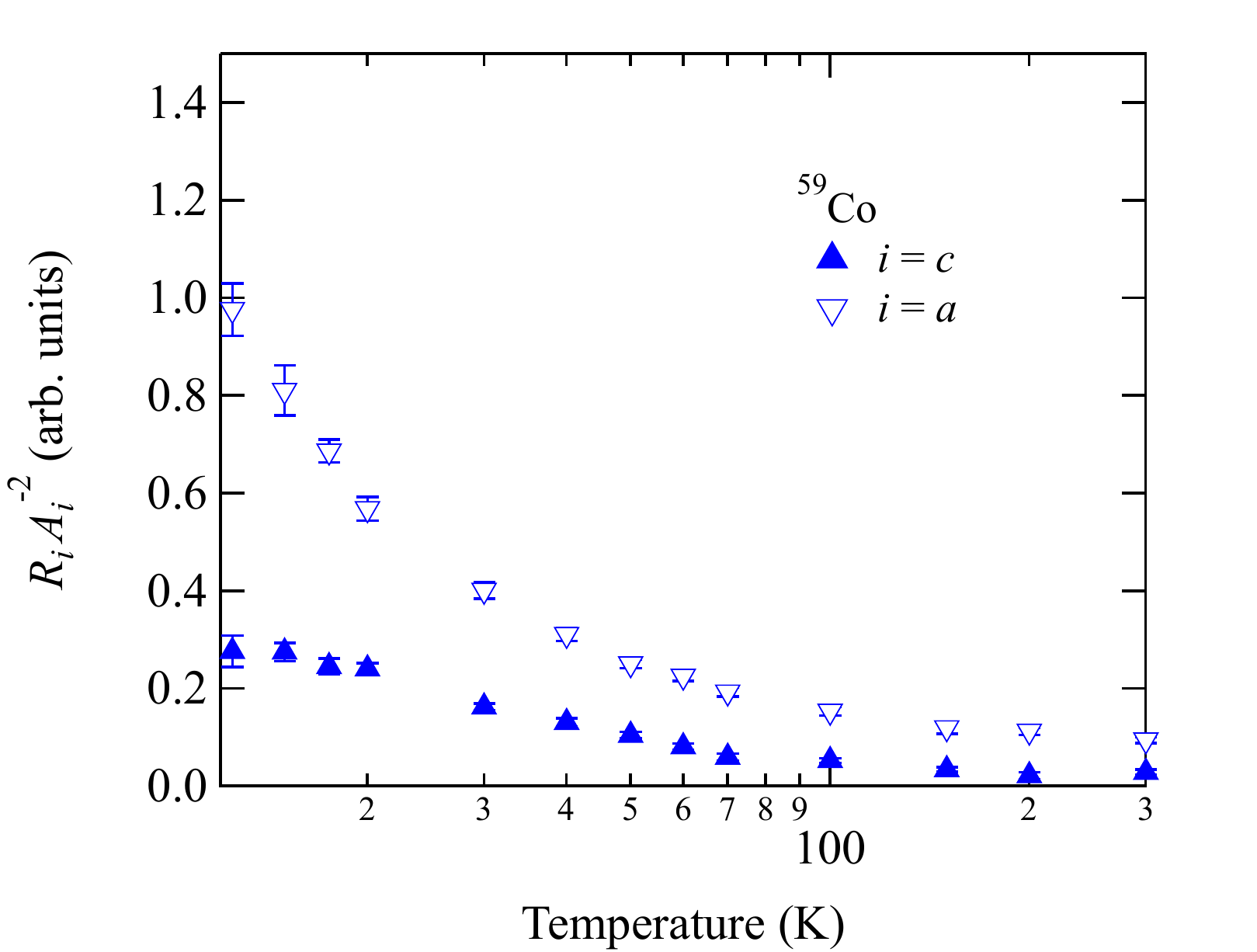}
	\caption{Hyperfine-coupling-normalized spin fluctuation rate $R_i/A_i^2$ versus temperature for ${}^{59}$Co.}
	\label{fig:Co59_SFaniso}
\end{figure}

$R_i/A_i^2$ versus temperature (shown in Fig.~\ref{fig:Co59_SFaniso}) reveals that the spin fluctuations are anisotropic at the ${}^{59}$Co site, but the anisotropic hyperfine coupling results in a nearly isotropic spin-lattice relaxation rate. This result is consistent with the upturn in the two lowest temperature points of $T_1^{-1}$ for $H_0\parallel c$ with respect to the corresponding data for $H_0\parallel a$. These more intense spin fluctuations in the basal plane drive faster relaxation for $H_0\parallel c$ at low temperatures.

\begin{figure}
	\includegraphics[width=\linewidth]{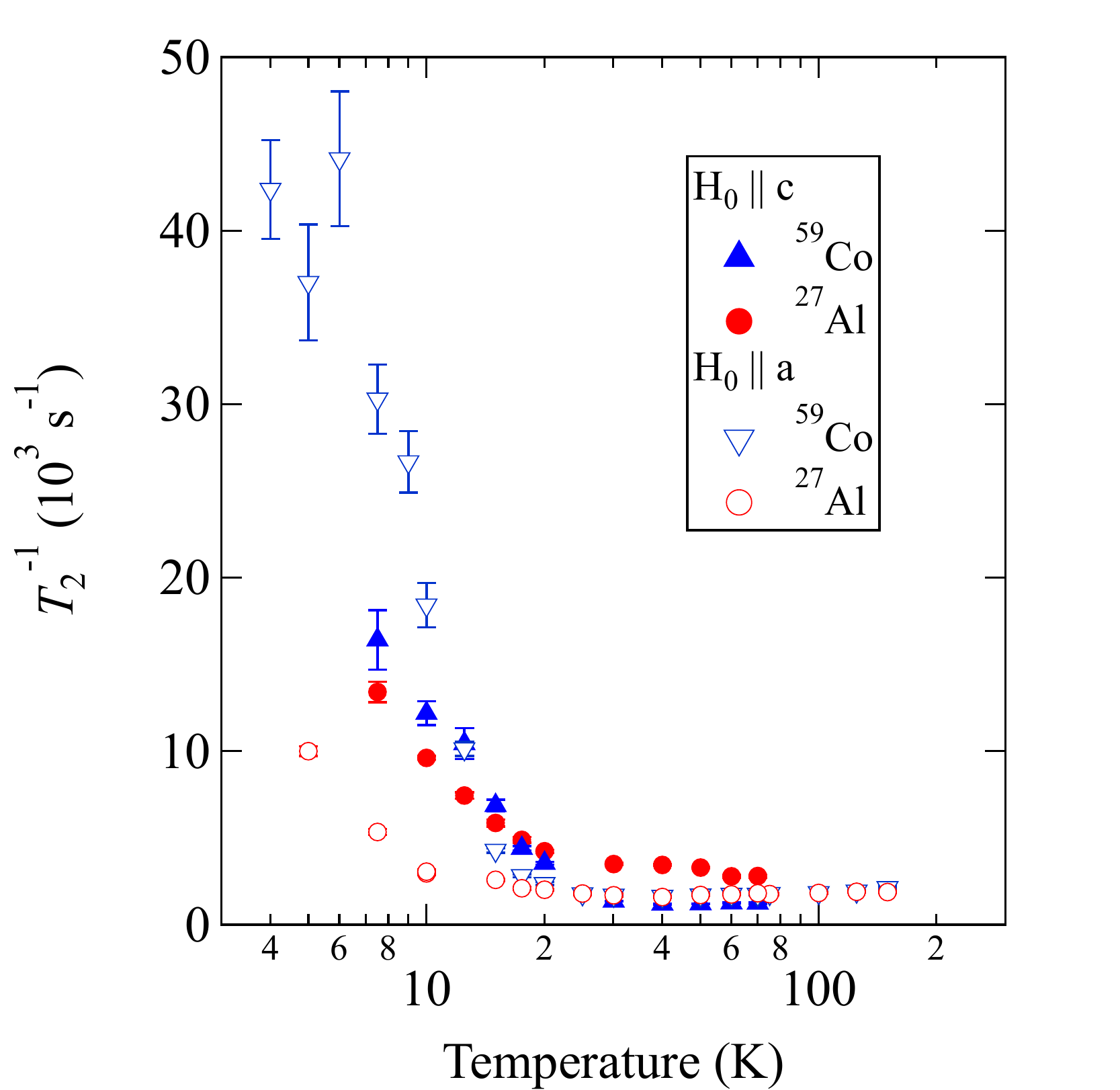}
	\caption{Spin-spin relaxation rate $T_2^{-1}$ of \ccag{} versus temperature.}
	\label{fig:T2}
\end{figure}

At low temperatures, but above the magnetic transition, the Co spectrum and parts of the Al spectrum become wiped out due to strong relaxation (see Figs.~\ref{fig:FS_spectra}~and~\ref{fig:HS_spectra}), such that we can no longer observe a spin echo. Indeed, upon measurement of the spin-spin relaxation rate $T_2^{-1}$ (see Fig.~\ref{fig:T2}), we find that it becomes extremely fast at low temperatures. To determine if we still observe the full ensemble of nuclei, we calculated the number of observed nuclei
\begin{equation}
\label{eqn:wipeout}
N_0 \propto \frac{T}{N_{avgs}(T)} \exp{\left(\frac{2\tau}{T_2(T)}\right)} \int S(H,T)dH.
\end{equation}
$N_{avgs}$ is the number of averaged spin echos, $\tau$ is the time between the pulses of the spin echo sequence over which the nuclear magnetization dephases, $T_2$ is the spin-spin relaxation time, and $S(H)$ is the spectral function, which we integrate over to access the spectral intensity. We find $N_0$ is constant within the propagated error for all temperatures, and therefore the spectral wipeout can be completely accounted for by fast $T_2^{-1}$ at low temperatures.

Furthermore, we note that at the ${}^{59}$Co site $T_2^{-1}$ becomes faster for $H_0\parallel a$. $T_2^{-1}$ is driven by fluctuations of the hyperfine field parallel to the applied external field, therefore this increase at low temperature is qualitatively consistent with the anisotropy of spin fluctuations extracted from $T_1^{-1}$ measurements discussed above. We show the results for ${}^{27}$Al for completeness, but purposely do not draw conclusions from the measured anisotropy of $T_2^{-1}$ at these sites due to the spectral overlap of the multiple sites.

\section{Conclusions}
\label{sec:conclusions}

We have investigated the NMR and NQR properties of the heavy fermion compound \ccag{} as a function of temperature. We also performed DFT calculations of the EFG at the ${}^{59}$Co and ${}^{27}$Al sites, which agree with the observed NMR and NQR spectra. At high temperatures, we find that the Knight shift is proportional to the bulk magnetic susceptibility and follows a Curie-Weiss temperature dependence. We observe a clear Knight shift anomaly in the $K$ versus $\chi$ plots indicating anisotropic coherence temperatures ${}^{59}T^*_c = 17.5$ K, ${}^{59}T^*_a = 10$ K, ${}^{27}T^*_a(3) = 12.5$ K that characterize the scale over which the heavy fermion nature of \ccag develops. We measured the spin-lattice relaxation rate and find that $(T_1T)^{-1} \propto T^{-1}$ for all measured sites and orientations at high temperature, indicating relaxation driven by $\mathbf{q}=0$ spin fluctuations of the local Ce moments. We find that the anisotropy of the spin fluctuations is significant and the fluctuations are larger in the basal plane than in the $c$ direction. Furthermore, the measured values of \slrr{} agree well with calculations of the expected value due to relaxation driven by local moment spin fluctuations. We find that $(T_1TK)^{-1} \propto T^{-1/2}$, which is expected for a Kondo system. In the low-temperature regime ($T \lesssim 3$ K), $T_1^{-1}$---measured by NQR at the ${}^{59}$Co $3\nu_Q$ transition---diverges indicating slowing down of spin fluctuations above the magnetic phase transition. Finally, the temperature dependence of $\chi T$ indicates weak ferromagnetic spin fluctuations above the magnetic transition.

\begin{acknowledgments}

We thank J. M. Lawrence, L. Civale, R. Movshovich, M. Janoschek, N. A. Wakeham, Y. Luo, D.-Y. Kim, D. Fobes, Z. Liu, N. Sung, N. Leon-Brito, A. M. Mounce, N. J. Curro, and H. Yasuoka for stimulating discussions. Work at Los Alamos National Laboratory was performed with support from two projects under the auspices of the U.S. Department of Energy, Office of Basic Energy Sciences, Division of Materials Sciences and Engineering. Magnetization measurements below 10 K were performed by the ``Towards a Universal Description of Vortex Matter in Superconductors'' project, while the remaining work was supported by the ``Complex Electronic Materials'' project. Part of the work was supported by the Los Alamos LDRD program. A. P. D. acknowledges a Director's Postdoctoral Fellowship supported through the Los Alamos LDRD program.
\end{acknowledgments}

\bibliography{Co-2174_NMR}

\end{document}